\documentclass[aps,prc,groupedaddress,twocolumn,showpacs,amsmath,amssymb]{revtex4}
\usepackage{graphics}
\usepackage{dcolumn}
\usepackage{longtable}

\begin{document}

\title{Analysis of uncertainties in $\alpha$-particle optical-potential assessment below the Coulomb barrier}
\author{V.~Avrigeanu} \email{vlad.avrigeanu@nipne.ro}
\author{M.~Avrigeanu}
\affiliation{Horia Hulubei National Institute for Physics and Nuclear Engineering, P.O. Box MG-6, 077125 Bucharest-Magurele, Romania}

\begin{abstract}
\noindent
{\bf Background:} 
Recent high-precision measurements of $\alpha$-induced reaction data below the Coulomb barrier have pointed out questions of the $\alpha$-particle optical-model potential (OMP) which are yet open within various mass ranges. 
{\bf Purpose:} 
The applicability of a previous optical potential and eventual uncertainties and/or systematic errors of the OMP assessment at low energies can be further considered on this basis. 
{\bf Method:} Nuclear model parameters  based on the analysis of recent independent data, particularly $\gamma$-ray strength functions, have been involved within statistical model calculation of the $(\alpha,x)$ reaction cross sections. 
{\bf Results:} 
The above-mentioned potential provides a consistent description of the recent $\alpha$-induced reaction data with no empirical rescaling factors of the $\gamma$ and/or nucleon widths. 
{\bf Conclusions}: 
A suitable assessment of $\alpha$-particle optical potential below the Coulomb barrier should involve the statistical-model parameters beyond this potential on the basis of a former analysis of independent data. 
\end{abstract}


\maketitle

\section{INTRODUCTION}
Recent high-precision measurements \cite{ps14,ggk15,cy15,ln15,ao15,as15} of $\alpha$-particle induced reaction data below the Coulomb barrier ($B$) provide an useful opportunity to investigate the results of a previous optical-model potential (OMP) for $\alpha$-particles on nuclei within the mass number range 45$\leq$$A$$\leq$209 \cite{va14}. Actually, this potential was established by (1) analysis of $\alpha$-particle elastic-scattering angular distributions above $B$ \cite{va14,ma03}, and (2) Hauser-Feshbach statistical model (SM) assessment of the available $(\alpha,\gamma)$, $(\alpha,n)$ and $(\alpha,p)$ reaction cross sections for incident energies below $B$ and target nuclei either with $A$$\leq$120 \cite{ma09} or heavier \cite{va14,ma10}. Thus, starting from a semi-microscopic OMP with a double-folding model (DFM) real part and a phenomenological energy-dependent imaginary-potential dispersive contribution \cite{ma03,ma09}, a full phenomenological analysis of the same data led to a spherical OMP to be easily involved within SM calculations for basic and applied objectives. Main points of this potential are (i) a strongly-modified energy dependence of the surface imaginary-potential depth below 0.9$B$ \cite{ma09} which is now a reference energy in this respect (e.g., \cite{as11}), and (ii) an energy-dependent radius of the surface imaginary part for the well-deformed nuclei \cite{va14}. 

A consistent description of all $\alpha$-induced reaction data available at that time \cite{ma09,ma10,va14} was provided using this potential and especially no empirical rescaling factors of the $\gamma$ and/or neutron widths. However, the above-mentioned recent works have pointed out $\alpha$-particle OMP questions  yet open within various mass ranges. Thus, a detailed systematic study of the $(\alpha,\gamma)$ reactions for all stable nickel isotopes \cite{as15}, following also recent distinct studies of this reaction on $^{58}$Ni \cite{sjq14} and $^{62}$Ni \cite{as07} isotopes, aimed to provide a constraint for the choice of input models in a given $A$ range. Nevertheless, different best combinations of input parameters for the TALYS 1.6 code \cite{TALYS} were found for each of the investigated isotopes while the combination identified that reproduces simultaneously the experimental data for all Ni isotopes has been at variance with the findings of the distinct studies \cite{sjq14,as07} as well as the  grounds of the concerned $\alpha$-particle OMP \cite{pd02}. Actually, the particular study of the $^{58}$Ni$(\alpha,\gamma)$$^{62}$Zn reaction had already mentioned that further theoretical work is required to obtain a full understanding of $\alpha$-induced reaction data on $^{58}$Ni \cite{sjq14}. 

The $(\alpha,n)$ and $(\alpha,\gamma)$ reaction cross sections of several proton-rich nuclei within the region just above the $A$$\approx$100 form also the object of several recent studies at low energies, with similar results for the reliability of SM predictions. So, it was possible to reproduce simultaneously the data for $^{107}$Ag only by rescaling the ratio of $\gamma$ to neutron widths \cite{cy15}. A study of both elastic-scattering and $\alpha$-induced reaction cross sections for $^{106}$Cd at low energies concluded as well that these data allow to constrain the other SM ingredients except the $\alpha$-particle OMP \cite{ao15}. Total and for the first-time measured partial cross sections of the $(\alpha,\gamma)$ reaction, for $^{112}$Sn target nucleus, pointed out a disagreement between experiment and theory which was considered to be most probably a deficiency of the nuclear-physics input \cite{ln15}. 

The $(\alpha,n)$ reaction cross-section studies performed in the meantime also for the heavier nuclei $^{164,166}$Er \cite{ggk15} and $^{187}$Re \cite{ps14}, with the main goal to test the low energy modification \cite{as11} of the widely-used $\alpha$-particle OMP of McFadden and Satchler \cite{mcf66}, concluded that the corresponding energy-dependence steepness assumed for a Fermi-type volume imaginary potential has the different parameter values of 2.5 and 4 MeV, respectively.
Moreover, it has been considered an yet open question \cite{ps14,ggk15,cy15,ao15} whether this modification is due to a required OMP change, which affects the total-reaction cross section, or due to the neglection of the Coulomb excitation (CE) in the entrance channel \cite{tr13}. 
All these new measured data and related $\alpha$-particle OMP studies within so different mass ranges motivated the present work, with the aim to both check the applicability of the previous OMP \cite{va14} and to look for uncertainties and possible systematic errors of $\alpha$-particle optical-potential assessment at low energies. 
 
Contrary to the conclusions of the above-mentioned studies of $\alpha$-particle induced reaction data at low energies, obtained on the basis of these data fit using a range of global input parameters for the SM calculations, we pay attention firstly to the use of a consistent input parameter set, either established or validated by analyzing various independent data. This set is briefly described in Sec.~\ref{SMcalc} of this work, while more detailed presentation was given previously \cite{va14,ma09b,va15}. The results corresponding to the OMP of Ref. \cite{va14} are then compared with the recently measured cross sections \cite{ps14,ggk15,cy15,ln15,ao15,as15} in Sec.~\ref{Res}, followed by conclusions in Sec.~\ref{Conc}.

\section{Nuclear models and parameters} \label{SMcalc}

\subsection{Coulomb excitation effects on $\alpha$-particle OMP setting up} 

The overestimation of the $\alpha$-induced reaction cross sections at low energies provided by the use of the $\alpha$-particle OMP of McFadden and Satchler \cite{mcf66} has been explained recently \cite{tr13} by the neglect of the CE process, i.e., the $\alpha$-particle inelastic scattering by the electric field of the target nucleus. 
Consequently, the CE diversion of the incident flux from the compound-nucleus (CN) channel was taken into account by decreasing the OMP transmission coefficients given for each partial wave. Because the corresponding renormalization of the $\alpha$-particle total-reaction cross section $\sigma_R$ does not affect the $\alpha$-particle emission, the CE effects in the incident channel have been considered at the origin of the difference between the OMPs corresponding to the incident and emitted $\alpha$-particles, respectively \cite{tr13}. 

However, following former comments on this assumption \cite{va14}, it has been shown that the corresponding partial waves and integration radii provide evidence for the distinct account of the CE cross section and OM $\sigma_R$ \cite{va16}. 
Thus the largest contribution to CE cross section comes by far from partial waves larger than the ones contributing to the $\sigma_R$ values. 
Therefore one should pay attention only to the assessment of direct-interaction (DI) collective $\alpha$-particle inelastic scattering cross section which should be subtracted from $\alpha$-particle $\sigma_R$ in order to obtain the corresponding CN cross section which is to be then involved in statistical model calculations. 
It should include the effects of the CE+nuclear interference that correspond to an integration radius which is typical to the short-range nuclear interactions, i.e., of $\sim$15 fm. On the other hand, the cross sections related to these effects are usually much lower than the $\sigma_R$ values even below $B$ \cite{va16}. 
    
\subsection{Statistical model parameters} 

\begingroup
\squeezetable
\begin{table*} 
\caption{\label{densp} Low-lying levels number $N_d$ up to excitation energy $E^*_d$ \protect\cite{ensdf} used in reaction cross-section SM calculations, the low-lying levels and $s$-wave nucleon-resonance spacings $D_0^{\it exp}$ (with uncertainties given in parentheses, in units of the last digit) in the energy range $\Delta$$E$ above the separation energy $S$, for the target-nucleus ground state (g.s.) spin $I_0$, fitted to obtain the BSFG level-density parameter {\it a} and g.s. shift $\Delta$ (for a spin cutoff factor calculated with a variable moment of inertia \cite{va02} between half and 75\% of the rigid-body value, from g.s. to $S$, and reduced radius $r_0$=1.25 fm), and the average radiation widths $\Gamma_{\gamma}$, either measured \cite{ripl3} or based on systematics (given between square brackets), and corresponding to the EGLO model parameter $T_f$=0.7 MeV used for description of the RSF data of the Zn isotopes \cite{kn82,kn80,ripl2,be79,be80,kn83,as14,mg14}.}
\begin{ruledtabular}
\begin{tabular}{cccccccrccrc}
Nucleus &$N_d$&$E^*_d$& \multicolumn{6}{c}
  {Fitted low-lying levels and nucleon-resonance data}& $a$ & $\Delta$\hspace*{3mm}& $\Gamma_{\gamma}$ \\
\cline{4-9}
  &  &    &$N_d$&$E^*_d$&$S+\frac{\Delta E}{2}$&$I_0$& $D_0^{\it exp}$ & $\Gamma_{\gamma}$ \\ 
          &  &(MeV)&  &(MeV)& (MeV) &   &   (keV)  & (meV) &(MeV$^{-1}$)&(MeV)& (meV) \\ 
\hline
$^{58}$Ni&34&4.574&34&4.574&      &   &             &          &5.90& 0.40 \\ 
$^{60}$Ni&50&4.579&51&4.613&      &   &             &          &6.00& 0.06 \\ 
$^{61}$Ni&35&2.863&35&2.863& 8.047& 0 & 13.8(9)$^a$ &1120(200) &6.55&-0.93 \\ 
$^{62}$Ni&46&4.455&46&4.455&10.631&3/2& 2.10(15)$^a$&2000(500) &6.36& 0.27 \\ 
$^{64}$Ni&20&3.849&51&4.613&      &   &             &          &6.90& 0.75 \\ 
$^{61}$Cu&36&3.042&38&3.092&      &   &             &          &6.75&-0.64 \\
$^{63}$Cu&60&3.291&105&3.92& 9.026& 0 &  5.9(7)$^a$ &          &7.00&-0.65 \\ 
$^{64}$Cu&40&1.780&40&1.780& 7.993&3/2& 0.70(9)$^b$ & 490(30)  &7.70&-1.55 \\ 
$^{65}$Cu&40&3.132&48&3.278&      &   &             &          &7.85&-0.10 \\ 
$^{66}$Cu&22&1.439&22&1.439& 7.116&3/2& 1.30(11)$^a$& 385(20)  &7.88&-1.40 \\ 
$^{67}$Cu& 6&1.937& 5&1.670&      &   &             &          &8.20& 0.07 \\ 
$^{61}$Zn&15&1.730&15&1.730&      &   &             &          &6.70&-1.25 \\ 
$^{62}$Zn&32&4.217&32&4.217&12.890&3/2&             &[730(330)]&6.20& 0.16 & 1380 \\ 
$^{63}$Zn&21&1.978&20&1.909&      &   &             &          &7.50&-0.93 \\ 
$^{64}$Zn&60&3.993&75&4.319&11.862&3/2&             &[510(230)]&6.80&-0.04 & 740 \\ 
$^{65}$Zn&35&2.216&35&2.216& 8.018& 0 &  2.3(3)$^a$ & 726(60)  &8.33&-0.71 & 200 \\
$^{66}$Zn&33&3.738&39&3.825&11.059&5/2&             &[450(180)]&7.70& 0.55 & 290 \\
$^{67}$Zn&40&2.175&40&2.175& 7.278& 0 & 4.62(55)$^a$& 390(60)  &8.11&-0.97 \\ 
$^{68}$Zn&51&3.943&51&3.943&10.291&5/2& 0.37(2)$^a$ & 440(60)  &8.05& 0.59 & 460 \\ 
$^{107}$Ag&19&1.325&19&1.325&      &   &            &          &13.70&-0.23 \\ 
$^{106}$Cd&22&2.566&24&2.630&      &   &            &          &13.00& 0.86 \\
$^{110}$Cd&36&2.662&38&2.707&      &   &            &          &13.75& 0.82 \\
$^{109}$In&17&1.816&16&1.759&      &   &            &          &14.00& 0.29 \\ 
$^{110}$In&25&1.006&30&1.063&      &   &            &          &14.50&-0.61 \\
$^{111}$In&36&2.112&36&2.112& 9.992&7.0&            &[180(60)] &14.40& 0.35 & 125 \\
$^{109}$Sn&13&1.269&13&1.269&      &   &            &          &13.90&-0.12 \\ 
$^{110}$Sn&30&3.183&30&3.183&11.270&5/2&            &[180(90)] &13.40& 1.35 & 115 \\ 
$^{112}$Sn&23&2.986&23&2.986&      &   &            &          &13.85& 1.34 \\  
$^{115}$Sb&15&2.104&15&2.104&      &   &            &          &14.00& 0.65 \\ 
$^{115}$Te& 8&1.272& 8&1.272&      &   &            &          &14.00& 0.09 \\ 
$^{116}$Te& 8&1.812& 8&1.812&11.278&7/2&            &[295(20)] &14.00& 0.63 & 290 \\
$^{164}$Er&39&1.766&39&1.766&      &   &            &          &17.00& 0.16 \\
$^{167}$Tm&26&0.631&26&0.631&      &   &            &          &18.20&-0.77 \\ 
$^{167}$Yb&30&0.571&34&0.628&      &   &            &          &18.50&-0.83 \\ 
$^{168}$Yb&28&1.551&28&1.551& 9.062&5/2&            &[110(30)] &17.50& 0.11 & 150 \\
\end{tabular}	 
\end{ruledtabular}
\begin{flushleft}
$^a$Reference \cite{ripl3}\\
$^b$Reference \cite{hv88}\\
\end{flushleft}
\end{table*}
\endgroup

We have also used within the present $(\alpha,x)$ reaction analysis a consistent set of (i) nucleon \cite{KD03} and (ii) $\gamma$-ray transmission coefficients, and (iii) back-shifted Fermi gas (BSFG) nuclear level densities \cite{hv88}. They have been established or validated on the basis of independent measured data for neutron total cross sections and $(p,n)$ reaction cross sections \cite{exfor}, $\gamma$-ray strength functions \cite{kn82,kn80,ripl2,be79,be80,kn83,as14,acl13,hkt11,mg14} and $(p,\gamma)$ reaction cross sections \cite{exfor}, and low-lying levels \cite{ensdf} and resonance data \cite{ripl3}, respectively. 
Only the points in addition to the details given formerly \cite{va14,ma09b,va15} as well as the particular parameter values are mentioned hereafter.

The SM calculations discussed in the following were carried out mostly within a local approach using an updated version of the computer code STAPRE-H95 \cite{ma95}, with $\sim$0.2 MeV equidistant binning for the excitation energy grid.
The latest version 1.8 of the code TALYS \cite{TALYS}, with the OMP \cite{va14} stated as the default option for $\alpha$-particles, has also been used in the particular case of the $^{187}$Re$(\alpha,n)$$^{190}$Ir reaction. The presently calculated reaction cross sections have been compared with the evaluated data within the TENDL-2015 library \cite{TENDL15}, too, for an overall excitation-function survey. Unfortunately, the latter data are available over an 1-MeV equidistant energy grid, for the incident energies within this work, which is less usefully just above the reaction thresholds. 

\subsubsection{Nuclear level densities} 

The BSFG model parameters used in the following SM calculations are given in Table~\ref{densp}, following the low-lying levels numbers and excitation energies \protect\cite{ensdf} used in the SM calculations (the 2nd and 3rd columns), as well as the independent data that have been involved in their setting up for the nuclei with $A$$\sim$60, $A$$\sim$110, and $A$$\sim$160, in the case of updates or in addition to the parameters and data given formerly within Refs. \cite{va15}, \cite{ma09b}, and \cite{va14}, respectively. 

\subsubsection{Optical model potentials} \label{OMP}

\begin{figure*} 
\resizebox{1.5\columnwidth}{!}{\includegraphics{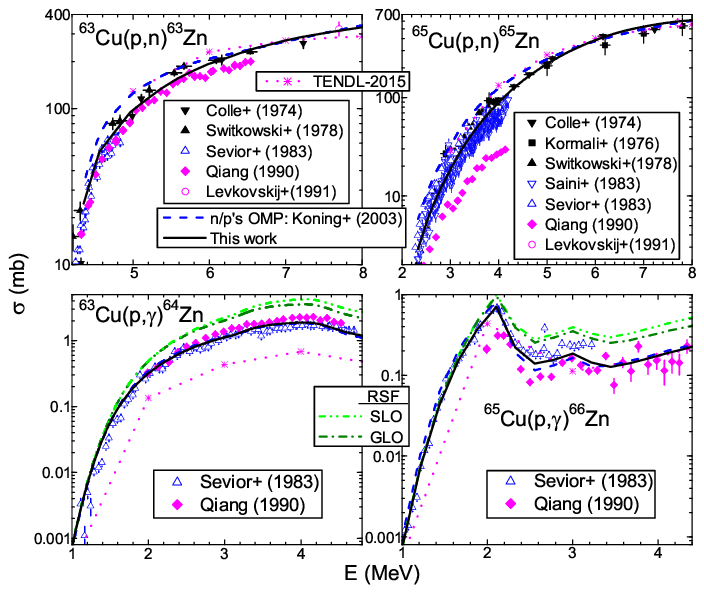}}
\caption{\label{Fig:Cu635pgn}(Color online) Comparison of the $(p,n)$ and $(p,\gamma)$ reaction cross sections for $^{63,65}$Cu target nuclei: measured \cite{exfor}, evaluated within the TENDL-2015 library \cite{TENDL15} (asterisks and dotted lines in-between), and calculated using the proton OMP parameters of either the global set \cite{KD03} (dashed curves) or finally adopted (solid curves), with addition in the latter case of the results for the $(p,\gamma)$ reaction corresponding also to the SLO (dash-dot-dotted curves) and GLO (dash-dotted curves) RSF models.}
\end{figure*}

{\it The neutron optical potential} of Koning and Delaroche \cite{KD03} was obviously the first option, paying however the due attention to the authors' remark that their global potential does not reproduce the minimum around the neutron energy of 1-2 MeV for the neutron total cross sections $\sigma_T$ of the $A$$\sim$60 nuclei. In spite of the scarce $\sigma_T$ data base available for the Zn isotopes  \cite{exfor}, we found that the global potential \cite{KD03} provides a suitable description of these data for the neutron energies starting from 2 MeV, while there is an $\sim$40\% overestimation for the energies below 1 MeV. Because the latter energy range is most important for the statistical CN de-excitation within the $(\alpha,x)$ reactions below $B$, we looked for a better description of this energy range. Thus we found that the calculated $\sigma_T$ values by using the OMP of Engelbrecht and Fiedeldey \cite{cae67} are in good agreement with the experimental data. Therefore we involved this neutron OMP in the SM calculations for Ni isotopes. 

{\it The proton optical potential} of Koning and Delaroche \cite{KD03} was also firstly considered for the calculation of the proton transmission coefficients, too. A particular check was found however desirable for the residual Cu isotopes because a previous study showed that the measured proton total reaction cross sections $\sigma_R$ available for these isotopes at energies $\leq$10 MeV \cite{rfc96} are overestimated (Fig. 1 of Ref. \cite{ma08}). Therefore, an analysis of the $(p,n)$ reaction cross sections was carried out within several MeV above the $(p,n)$ reaction effective thresholds for $^{63,65}$Cu target nuclei,  with the results shown in Fig.~\ref{Fig:Cu635pgn}. 

The corresponding SM calculations were obviously carried out using the same input parameters as in the rest of this work and the similar one on Zn isotopes \cite{va15}, except the DI inelastic-scattering cross sections of only several percents for the proton energies up to 8 MeV, to be taken into account for the decrease of the $\sigma_R$ finally involved within CN calculations. 
The $\sim$40\% overestimation of the measured data \cite{exfor} especially in the first 2-3 MeV of these excitation functions is visible also for the evaluated cross sections of the TENDL-2015 library \cite{TENDL15}. This overestimation has been removed (Fig.~\ref{Fig:Cu635pgn}) by using in the present work a local OMP for the low-energy protons on the Cu isotopes \cite{ss83}.

{\it The $\alpha$-particle optical potential} for nuclei within the 45$\leq$$A$$\leq$209 range \cite{va14} has been used for both $\alpha$-induced reaction and $\alpha$-emission calculations, following the conclusions corresponding to the $A$$\sim$60 nuclei \cite{va15}.

The same OMP could be also involved in DI distorted-wave Born approximation (DWBA) calculation of the cross sections for the collective inelastic scattering, using the corresponding deformation parameters of the first 2$^+$ and 3$^-$ collective states, and the CE collective form \cite{va15,va16}. However, typical DI inelastic-scattering cross sections, taken into account for the decrease of the $\alpha$-particle $\sigma_R$ within the CN calculations for, e.g., the target nucleus $^{60}$Ni grow up from $\sim$1\% to $<$6\% of $\sigma_R$ for $\alpha$-particle energies from 5 to 7.6 MeV, and then decrease.

\subsubsection{$\gamma$-ray strength functions} 

\begin{figure} 
\resizebox{0.75\columnwidth}{!}{\includegraphics{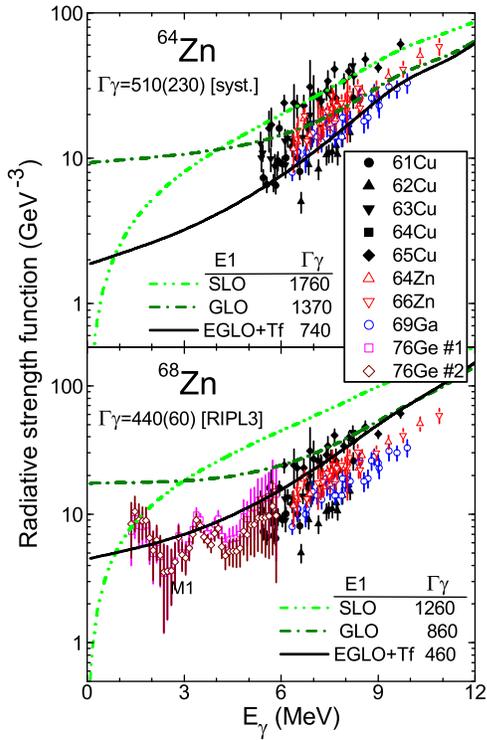}}
\caption{\label{Fig:RSF_Zn6468}(Color online) Comparison of the sum of calculated $\gamma$-ray strength functions of the $E1$ and $M1$ radiations for $^{64}$Zn and $^{68}$Zn nuclei,  using to the SLO (dash-dot-dotted curves), GLO (dash-dotted curves), and EGLO (solid curves) models for E1 radiations, and SLO model for M1 radiations, as well as of the calculated average $s$-wave radiation widths $\Gamma_{\gamma}$ (in meV). There are also shown the measured dipole $\gamma$-ray strength functions for $^{61,62,63,64,65}$Cu, $^{64,66}$Zn, $^{69}$Ga and $^{76}$Ge nuclei \cite{kn82,kn80,ripl2,be79,be80,kn83,as14}, and the $\Gamma_{\gamma}$ values either measured or deduced from systematics of the measured data \cite{ripl3}.} 
\end{figure}

\begin{figure} 
\resizebox{1.0\columnwidth}{!}{\includegraphics{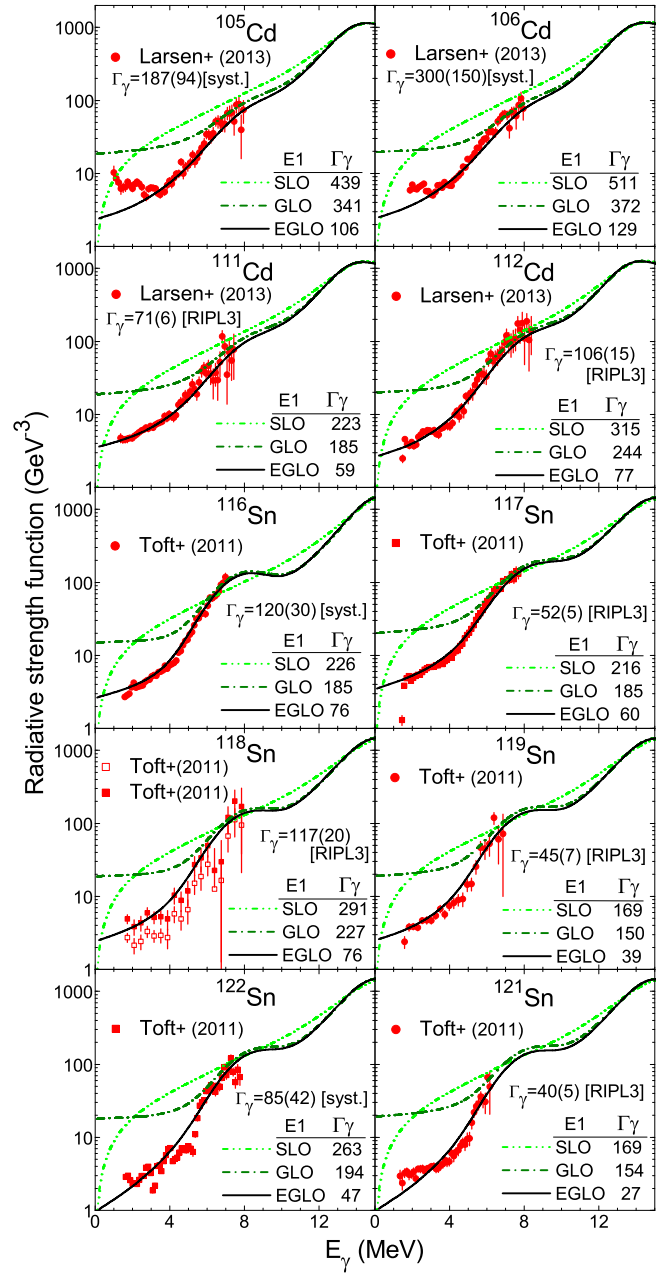}}
\caption{\label{Fig:RSF_CdSn}(Color online) As Fig.~\ref{Fig:RSF_Zn6468} but for $^{105,106,111,112}$Cd \cite{acl13} and $^{116-119,121,122}$Sn \cite{hkt11} nuclei.}
\end{figure}

Contrary to the use of $\gamma$-ray strength functions established by a renormalization carried out in order to achieve agreement with the $(\alpha,\gamma)$ data (e.g., Ref. \cite{cy15}), we rely on the measured data of radiative strength function (RSF) and average $s$-wave radiation widths $\Gamma_{\gamma}$ \cite{ripl3} (Table~\ref{densp}). There have been used in this respect the formerly measured RSFs \cite{kn82,kn80,ripl2,be79,be80,kn83} and one high-accuracy measurement at lower energies \cite{as14} in the neighborhood of the compound nuclei around $A$$\sim$60, and the recent systematic analyzes for Cd \cite{acl13} and Sn \cite{hkt11} isotopes, around $A$$\sim$110, while the similar data for heavier nuclei \cite{mg14} were already considered previously \cite{va14}. 

Moreover, the electric-dipole $\gamma$-ray strength functions, of main importance for calculation of the $\gamma$-ray  transmission coefficients, have been described by using the former Lorentzian (SLO) \cite{pa62}, generalized Lorentzian (GLO) \cite{jk91}, and enhanced generalized Lorentzian (EGLO) \cite{jk93} models. A constant nuclear temperature $T_f$ of the final states \cite{acl10} has been particularly used within the EGLO model. The giant dipole resonance (GDR) line-shape usual parameters have been derived from photoabsorption data, while the SLO model was used for the M1 radiation, with the global parametrization \cite{ripl3} for the GDR energy and width, i.e. $E_0$=41/$A^{1/3}$ MeV and $\Gamma_0$=4 MeV. 
Further details on particular parameter values are given hereafter. 

Comparison of the measured and calculated sum of $\gamma$-ray strength functions of the $E1$ and $M1$ radiations for the nuclei of interest within present work with $A$$\sim$60 (Fig.~\ref{Fig:RSF_Zn6468}) and $A$$\sim$110 (Fig.~\ref{Fig:RSF_CdSn}), similarly to   Fig. 3 of Ref. \cite{va14} for $A$$\sim$160, proves that both the SLO and GLO models overestimate the RSF data below the nucleon binding energy.  
The calculated $\Gamma_{\gamma}$ values corresponding to the adopted $\gamma$-ray strength functions are also shown in Figs.~\ref{Fig:RSF_Zn6468} and \ref{Fig:RSF_CdSn} for each E1 model involved in the present work. They are compared to the values either measured or deduced from systematics of the measured-data dependence on the neutron separation energy $S$ (Table~\ref{densp}). 
In spite of the low accuracy of these estimations for $A$$\sim$60 nuclei, it results also that only the EGLO $\gamma$-ray strength functions may provide values closer to the measured data. Moreover, just the same functions show a rather constant nonzero limit which is comparable to an average of the recent RSF data obtained for $^{74,76}$Ge nuclei \cite{as14}.

{\it $(p,\gamma)$ reaction data analysis} for $^{63,65}$Cu isotopes (Fig.~\ref{Fig:Cu635pgn}) has additionally been used to check the accuracy of the adopted RSFs for $A$$\sim$60 nuclei. Actually, the RSF uncertainties corresponding even to the EGLO model are largest for this mass range, in the absence of any detailed GDR and RSF recent study as in the case of the heavier nuclei. Thus, the GDR parameters derived from photoabsorption data for $^{64}$Zn \cite{be80} and $^{70}$Ge \cite{ripl3gamma} have been used for $^{62,64,65,66}$Zn and $^{68}$Zn excited nuclei, respectively. Finally, we adopted a $T_f$=0.7 MeV parameter value and a GDR peak cross section of 1.5 mb for the M1 radiation in order to obtain a compromise between the account of the RSF and $\Gamma_{\gamma}$ data as well (Fig.~\ref{Fig:RSF_Zn6468}). 

In spite of this rather rough EGLO approach, the comparison of the experimental and calculated $(p,\gamma)$ reaction data for $^{63,65}$Cu pointed out, however, a good agreement for the use of this option. While the measured data are well described by the parameters adopted in the present work, one may note a calculated cross-section increase of even a factor $>$2 for the incident energies above the $(p,n)$ reaction threshold if the EGLO model is replaced by the GLO one. An slightly larger factor is given by the use of the SLO model, which shows particularly that the RSF values at $\gamma$-energies larger than $\sim$3 MeV are more important for the calculation of capture cross sections. Thus it results that an eventual low-energy RSF enhancement \cite{as14}, if it exists for any of the Zn isotopes, would not affect essentially the calculated $(p,\gamma)$ reaction cross sections. 

\begin{figure*} 
\resizebox{2.0\columnwidth}{!}{\includegraphics{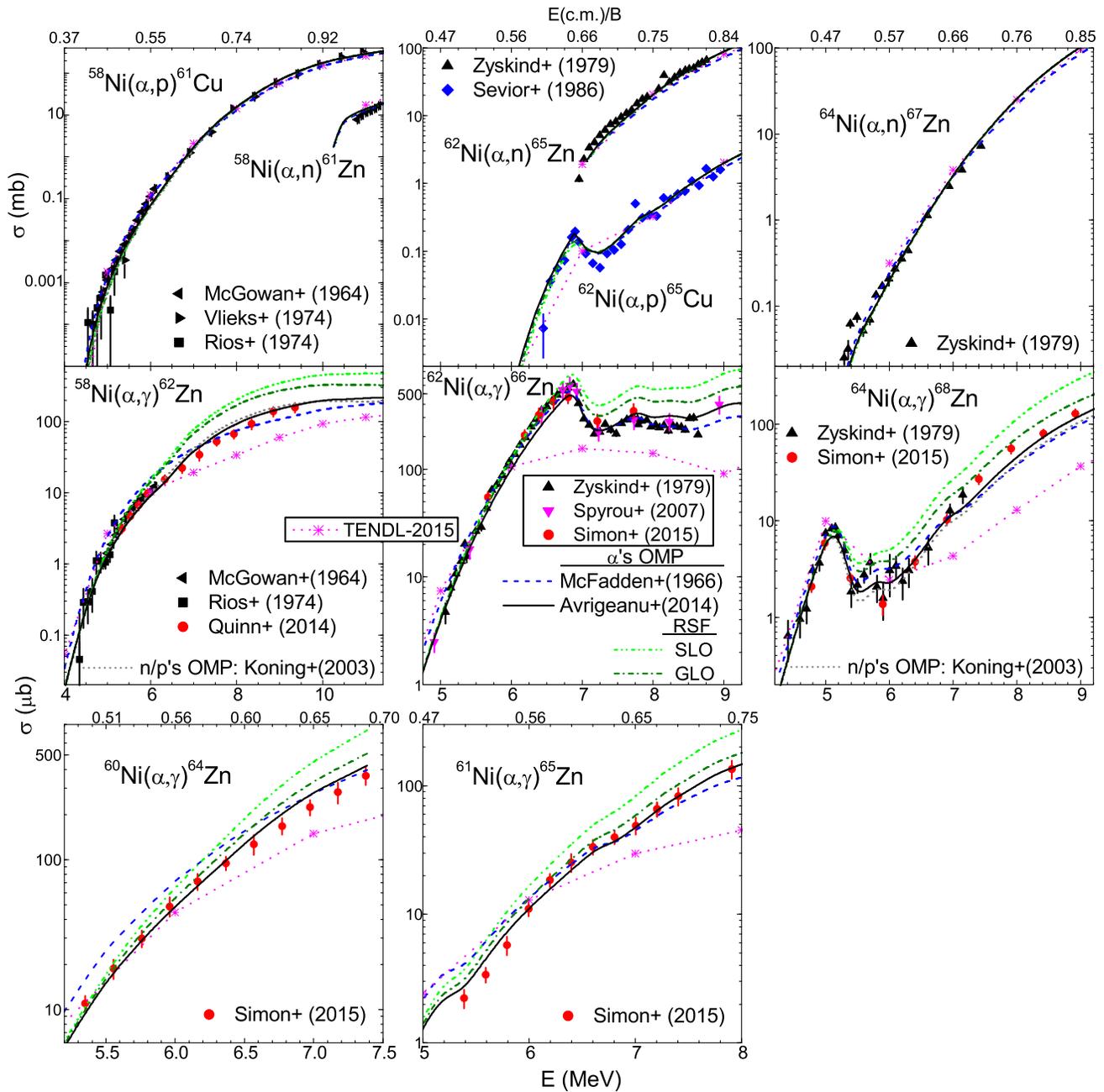}}
\caption{\label{Fig:Ni80124ax}(Color online) Comparison of the $\alpha$-induced reaction cross sections for $^{58,60-62,64}$Ni target nuclei: measured \cite{as15,exfor}, evaluated within the TENDL-2015 library \cite{TENDL15} (asterisks and dotted lines in-between), and calculated using the $\alpha$-particle global OMPs of either Ref. \cite{mcf66} (dashed curves) or Ref. \cite{va14} (solid curves), and the alternate involvements for the latter OMP of either the GLO (dash-dotted curves) and SLO (dash-dot-dotted curves) RSF models, or the global nucleon OMPs \cite{KD03} (short-dotted curves). }
\end{figure*}

\section{Results and Discussion} \label{Res}
\subsection{$(\alpha,x)$ reactions for stable Ni isotopes} 

The former version \cite{ma09} of the actual $\alpha$-particle potential \cite{va14} provided already a suitable description at once of all $(\alpha,\gamma)$, $(\alpha,n)$, and $(\alpha,p)$ reaction data available at that time for $^{58,62,64}$Ni and energies below $\sim$0.6$B$, 0.8$B$, and 0.7$B$, respectively. The new high-precision data \cite{as15,as07,sjq14} are particularly worthwhile for the present work as they (i) enlarge the corresponding energy range even above 0.8$B$ for the three above-mentioned isotopes, (ii) provide the firstly measured $(\alpha,\gamma)$ reaction cross section for $^{60,61}$Ni, and (iii) make possible a complete view of the $(\alpha,x)$ reaction cross sections below $B$ for a significant chain of stable isotopes (Fig.~\ref{Fig:Ni80124ax}). 

Because similar if not identical consistent input-parameter sets have formerly as well as presently been involved, the rather good agreement of all measured and calculated data shown in Fig.~\ref{Fig:Ni80124ax} is not surprising. The more important replacement of an early E1 model used in the past \cite{ma09} has little effect on the calculated cross sections as the corresponding former predictions were also checked versus the RSF and $\Gamma_{\gamma}$ data. Nevertheless, one may consider that the presently improved approach has led to the better description of the excitation-function structure above the $(\alpha,n)$ reaction threshold for $^{62,64}$Ni target nuclei.

On the other hand, the above comments concerning the RSF effects on the calculated $(p,\gamma)$ reaction cross sections are suitable for these  $(\alpha,\gamma)$ reactions as well, except the differences between the values obtained by using the various models. Thus, the use of the GLO model provides values larger by $\sim$50\% than the results related to the EGLO model, while a rather similar increase is provided in addition by the SLO model. This trend could be related to a slightly different $\gamma$-ray range involved within the decay of the same CNs populated by either protons or $\alpha$-particles, all above the energy of 3-4 MeV (below which the RSFs given by the SLO model are more and more lower than the GLO values). It results that the region just above this energy is involved in the case of the $(p,\gamma)$ reactions while higher energies but yet below the nucleon binding energies play a more significant role for the $(\alpha,\gamma)$ reactions. The most important point for the present work is still the consequent conclusion that an eventual low-energy RSF enhancement \cite{as14} would not affect essentially the calculated $(\alpha,\gamma)$ reaction cross sections. 

The systematic study of the $(\alpha,\gamma)$ reactions for stable Ni isotopes \cite{as15} makes possible a more sensitive comparison of the results corresponding to the global $\alpha$-particle potentials of Refs. \cite{va14,mcf66} as well as the advantage provided by this study versus that of the much-larger $(\alpha,p)$ and $(\alpha,n)$ reaction cross sections. Thus, the calculated results of the $(\alpha,\gamma)$ reaction definitely show cross sections increased by (50-100)\% at the lowest incident energies but (10-20)\% lower above 9 MeV, corresponding to a lower slope of the excitation function given by the potential of McFadden and Satchler \cite{mcf66}. 
This latter potential definitely works very well for an OMP having only four constant parameters and no surface imaginary component, except however the lowest energies 
 of largest interest for nuclear astrophysics and fusion technology.

On the other hand, the accuracy of these calculated $(\alpha,\gamma)$ reaction cross sections is not affected by an eventual less suitable knowledge of the nucleon OMPs. This has been found by replacing the above-mention OMPs by the global ones of of Koning and Delaroche \cite{KD03}, for the proton- and neutron-reach $^{58,64}$Ni target nuclei, with the $(\alpha,p)$ and $(\alpha,n)$ reaction cross sections being the largest fraction of the total reaction cross section, respectively. In spite of the above-mentioned (Sec.~\ref{OMP}) large differences between the nucleon OMPs adopted within this work for the Zn and Cu target nuclei, and Ref. \cite{KD03}, the $(\alpha,\gamma)$ reaction cross sections have been changed by less than 15\% through the use of the alternate OMP. A larger decrease but yet up to 23\% was obtained only around the minimum of the $^{64}$Ni$(\alpha,\gamma)$$^{68}$Zn excitation function just above the $(\alpha,n)$ reaction threshold.

Concerning the relation of the recent studies of the $(\alpha,\gamma)$ reactions for Ni isotopes \cite{as15,as07,sjq14} and present conclusions on the $\alpha$-particle potentials and RSF models, one may note that the combination of input parameters for the TALYS 1.6 code \cite{TALYS} which was identified that reproduces simultaneously the experimental data for all Ni isotopes \cite{as15} includes a potential which is the simpler version of the $\alpha$-particle OMPs of Ref. \cite{pd02}. However, the more complete and physical one, which is the third potential of that work, was found only formerly \cite{as07} to be most suitable for $^{62}$Ni. Also, the RSF which was found to be most suitable for $^{60,62}$Ni isotopes belongs to the GLO model, at variance with the RSF data shown in the present work. It results,  therefore, that compensation effects of less suitable options for various SM parameters can not be avoided, contrary to the use of a consistent parameter set.

\subsection{$(\alpha,x)$ reaction data analysis for $A$$\sim$110 nuclei} 

The recent RSF data of the Cd and Sn nuclei and their sensible analysis \cite{acl13,hkt11} have allowed a notably improved study of the $(\alpha,\gamma)$ reaction cross sections on $A$$\sim$110 nuclei. Because it is commonly assumed a small difference  in the RSFs of neighboring isotopes, we have used GDR, pygmy resonance, and $T_f$  parameters within the EGLO model obtained for some of these nuclei, as it is mentioned below. Actually, the use of these new RSF approach and the updated level-density parameters (Table~\ref{densp}) are the changes within this work with reference to our previous analysis for $^{106}$Cd \cite{ma09} and $^{112}$Sn \cite{ma09b} using the former version \cite{ma09} of the actual potential.

\subsubsection{$^{107}$Ag} 

\begin{figure} 
\resizebox{0.75\columnwidth}{!}{\includegraphics{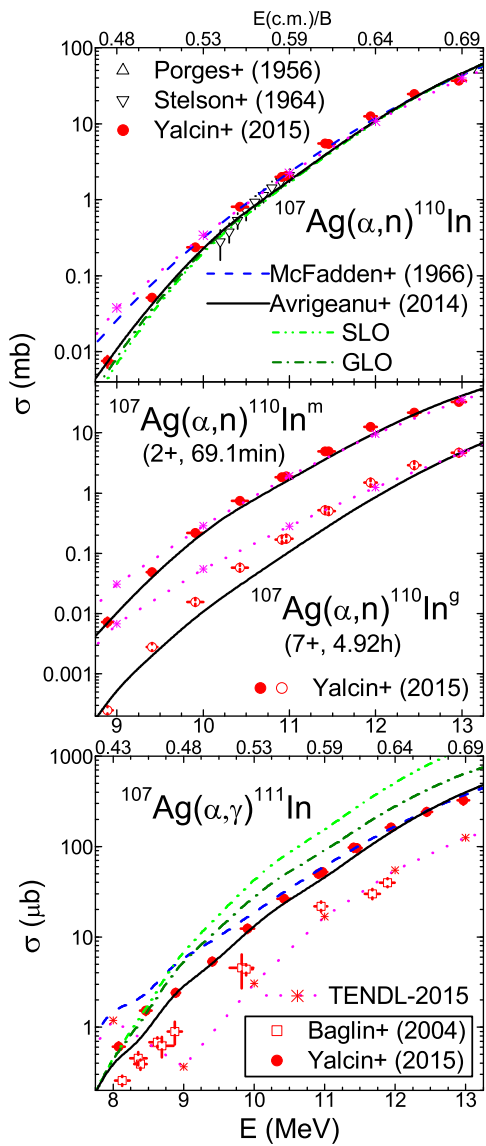}}
\caption{\label{Fig:Ag107ax}(Color online) As Fig.~\ref{Fig:Ni80124ax} but for $^{107}$Ag \cite{cy15,exfor,cmb05}.}
\end{figure}
 
The high-precision measured data for $^{107}$Ag \cite{cy15} have completed so usefully the earlier data available especially for the $(\alpha,n)$ reaction. On the other hand, a disagreement of these data for the $(\alpha,\gamma)$ reaction with a set only a decade older \cite{cmb05} made useful a further analysis as certain as possible. In this respect, we have considered that the use of the EGLO parameters \cite{acl13}, found to describe quite well the RSF data for $^{111}$Cd (see also Fig.~\ref{Fig:RSF_CdSn}), should be suitable also for $^{111}$In. An additional support for using these parameters comes from a very recent study of RSFs in the $^{114}$Cd$(\gamma,\gamma')$ and $^{113}$Cd$(n,\gamma)$ reactions \cite{rm16} which found similar experimental RSF values for $^{114}$Cd and $^{112}$Cd \cite{acl13} above 7 MeV. However, it has been found a difference at lower energies which was attributed also to the spin distribution of the states excited through the $(^3He,^3He')$ and $(\gamma,\gamma')$ processes. Therefore, the RSF data and parameters obtained by the Oslo group \cite{acl13,hkt11} are particularly suitable for the analysis of $(\alpha,\gamma)$ reactions. 
 
The general agreement between the present calculations and the measured data for this target nucleus (Fig.~\ref{Fig:Ag107ax}) has been obtained by using no rescaling factor or change of the $\alpha$-particle potential \cite{va14}. There is only a slight underestimation of the 7$^+$ ground state of the residual nucleus $^{111}$In $(\alpha,n)$ which could be also due to the knowledge of the corresponding decay scheme. On the other hand, a similar underestimation but for the lowest two measured points of the $(\alpha,\gamma)$ reaction may follow the use of a 0.2 MeV equidistant binning for the excitation energy grid. Beyond the validation of the $\alpha$-particle potential \cite{va14}, these results may additionally confirm the latest measured data set for the $(\alpha,\gamma)$ reaction \cite{cy15} with respect to the former one \cite{cmb05}.

\subsubsection{$^{106}$Cd} 

\begin{figure} 
\resizebox{0.75\columnwidth}{!}{\includegraphics{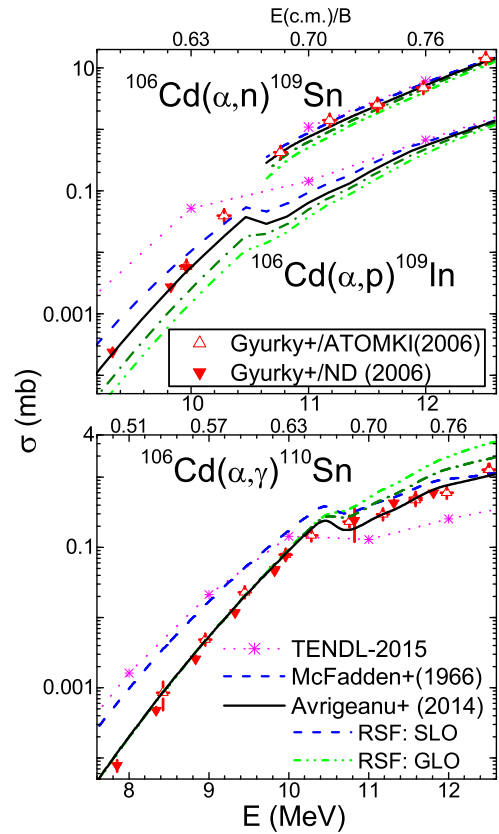}}
\caption{\label{Fig:Cd106ax}(Color online) As Fig.~\ref{Fig:Ni80124ax} but for $^{106}$Cd \cite{ao15,exfor}.}
\end{figure}

\begin{figure*} 
\resizebox{1.4\columnwidth}{!}{\includegraphics{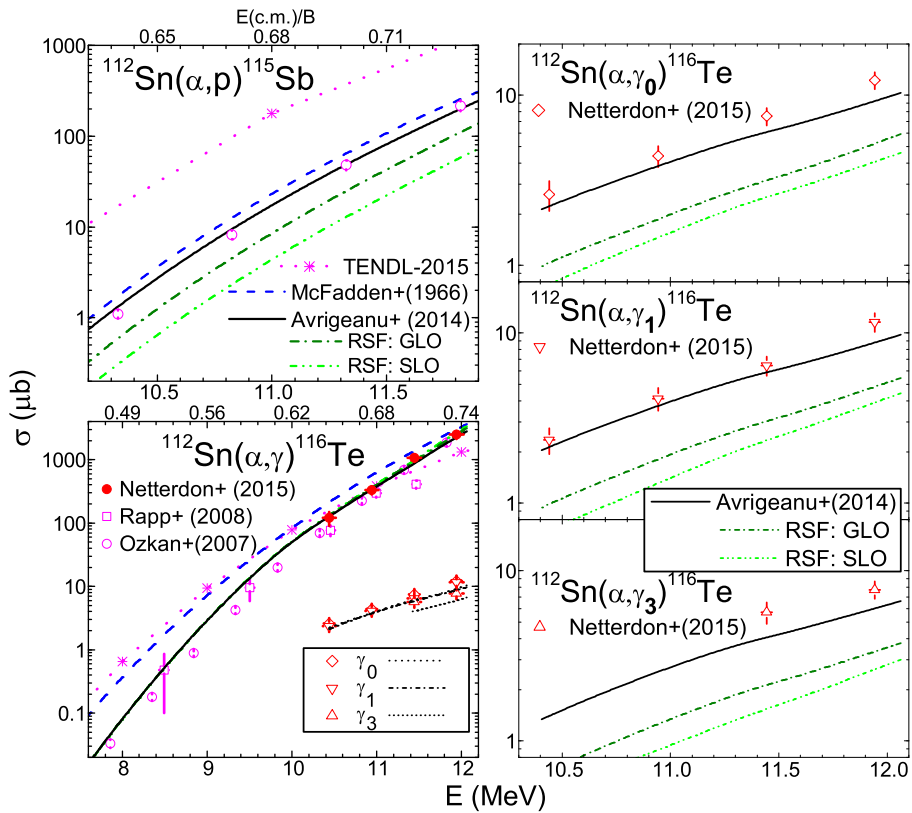}}
\caption{\label{Fig:Sn112ax}(Color online) As Fig.~\ref{Fig:Ni80124ax} but for $^{112}$Sn \cite{ln15,exfor} and the partial $(\alpha,\gamma)$ reaction cross sections for the decay of the entry state to the ground state, the first and the third excited states of the residual nucleus $^{116}$Te.}
\end{figure*}

A recent measurement of $\alpha$-particle elastic scattering angular distributions on $^{106}$Cd in the wide energy range from 16.1 to 27 MeV has been accompanied by a discussion of the $\alpha$-induced reaction cross sections below 12 MeV \cite{ao15}. A final remark of this work stated that conclusions on the $\alpha$-particle potential from the $\alpha$-induced reaction data are difficult in the case of this nucleus. The potential devoted particularly to heavy nuclei \cite{ma10} was used in the analysis of the elastic scattering analysis, on the correct premise that the latest version \cite{va14} would provide similar results. However, the former potential was used also within the reaction data analysis, where the same assertion became wrong. Nevertheless, mainly a large underestimation of the $(\alpha,p)$ reaction cross sections \cite{ao15}, related to the use of this potential, has triggered the following assay.

Actually, the same reaction data were already fully described by the former version of this potential (Fig. 15 of Ref. \cite{ma09}). 
The basic point of the present calculation has been the use of the RSF parameters of $^{106}$Cd \cite{acl13}, with the same above-mentioned motivation of close neighborhood of the corresponding compound nucleus $^{110}$Sn. 
The agreement of the calculated and measured data (Fig.~\ref{Fig:Cd106ax}) is again quite similar to that shown earlier \cite{ma09}, but now we point out the RSF effects on the calculated $(\alpha,\gamma)$ as well as $(\alpha,p)$ reaction cross sections. 
In fact, the negligible size of these effects for the dominant $(\alpha,\gamma)$ reaction below the particle-emission threshold stands as a major advantage of this reaction at similar energies for the $\alpha$-particle OMP study. 
On the other hand, at the same energies the minor $(\alpha,p)$ reaction cross sections decrease by factors of 2-3 and 3-5, respectively, when the EGLO model is replaced by the GLO and SLO options.
Then, above the $(\alpha,n)$ reaction threshold, the increase of the calculated $(\alpha,\gamma)$ reaction cross sections, by additional factors of $\sim$2 for the same changes, as well as a reducing decrease of the $(\alpha,p)$ reaction cross sections, follow naturally the corresponding RSFs. 
Therefore, the real cause of the $(\alpha,p)$ reaction data underestimation below and around  the $(\alpha,n)$ reaction threshold \cite{ao15} has been not the $\alpha$-particle potential but a deficient RSF involved within the SM analysis.

\subsubsection{$^{112}$Sn} 

A similar relation between the $(\alpha,\gamma)$ and $(\alpha,p)$ reaction cross sections, at incident energies below the $(\alpha,n)$ reaction threshold, there is also for the proton-rich too $^{112}$Sn target nucleus. The data measured for these reactions before 2010s were also suitably  described \cite{ma09,ma09b} while a new measurement of the $(\alpha,\gamma)$ reaction cross sections \cite{ln15} removes a minor discrepancy between two earlier data sets and actually has strengthened the validation of the present potential (Fig.~\ref{Fig:Sn112ax}). The RSF parameters of $^{112}$Cd \cite{acl13} have been used in this respect, with the following comment.

The analysis within Ref. \cite{ln15} had to assume several parameter adjustments \cite{ln15} including a reduction by a factor of 0.2 for the proton width provided by a default TALYS calculation, i.e., using the global nucleon optical potentials \cite{KD03}, in order to describe the data for $^{112}$Sn.
Then, a major underestimation of the $(\alpha,p)$ reaction cross section for $^{106}$Cd was obtained by using the same local parameters \cite{ln15}. 
This underestimation was obviously following the hard reduction factor of 0.2 adopted for the proton width while the global proton optical potential  \cite{KD03} was proved independently to work quite well in this mass region \cite{ma09b}. On the contrary, no scaling factor has been requested by the agreement shown in Fig.~\ref{Fig:Sn112ax} also for the $(\alpha,p)$ reaction. Thus, one may note the concurrent description of the $(\alpha,p)$ reaction data for both target nuclei $^{106}$Cd and $^{112}$Sn, with no change of the OMPs and/or RSF parameters. Moreover, even the changes of the calculated cross sections of this reaction, due to the use of alternate RSF models GLO and SLO, show nearly the same ratios (Fig.~\ref{Fig:Sn112ax}) as in the above case of $^{106}$Cd (Fig.~\ref{Fig:Cd106ax}).

Nevertheless, the main achievement of Ref. \cite{ln15}, namely the first measurement of partial $(\alpha,\gamma)$ cross sections, has made possible a really sound check of the adopted RSFs, beyond that of the $\alpha$-particle potential. The calculated partial cross sections shown in Fig.~\ref{Fig:Sn112ax} correspond only to the side feeding of the corresponding ground and excited states. The comparison of the experimental and calculated partial cross sections validated actually the whole SM approach as well as its two main ingredients for the present topic.

\subsection{$(\alpha,x)$ reaction data analysis for well-deformed nuclei} 

\begin{figure}  [t]
\resizebox{1.0\columnwidth}{!}{\includegraphics{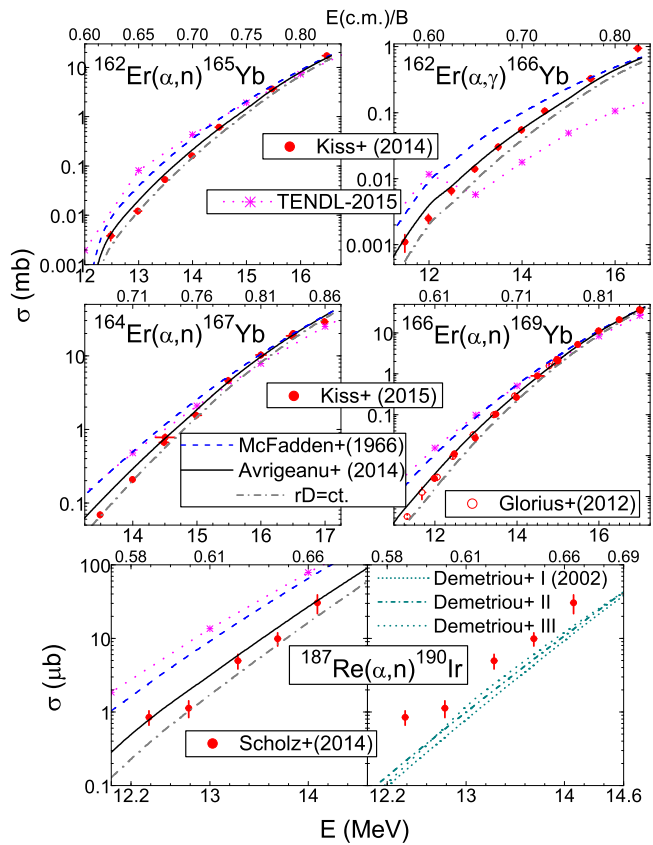}}
\caption{\label{Fig:Er16246Re} (Color online) Comparison of recently measured $(\alpha,x)$ reaction cross sections for $^{162}$Er \cite{ggk14}, $^{164,166}$Er \cite{ggk15,jg12}, and $^{187}$Re \cite{ps14}, with the evaluated data of the TENDL-2015 library \cite{TENDL15} (asterisks and dotted lines in-between) and calculated values using the $\alpha$-particle OMPs of Refs. \cite{mcf66} (dashed curves) and \cite{va14} with (solid curves) as well as without (dash-dotted curves) the energy-dependent radius for the lower-energy surface imaginary potential of the well-deformed nuclei. The $(\alpha,n)$ reaction cross sections calculated using the three versions of $\alpha$-particle OMP of Refs. \cite{pd02} (short dotted, short dash-dotted, and short dashed curves, respectively) are also shown for the target nucleus $^{187}$Re.}
\end{figure}

The $(\alpha,n)$ reaction cross-section studies performed in the meantime for the heavier nuclei $^{164,166}$Er \cite{ggk15} and $^{187}$Re \cite{ps14} made possible a further check of the previous inference of $\sim$7\% larger, at the lowest energies, as well as energy-dependent radius of the surface imaginary part of the $\alpha$-particle potential for the well-deformed nuclei \cite{va14}. 

\subsubsection{$^{164,166}$Er} 

The more recent $(\alpha,n)$ reaction data for $^{164,166}$Er \cite{ggk15} are compared with the calculation results and the former data for all Er isotopes \cite{ggk14,jg12} as well (Fig.~\ref{Fig:Er16246Re}), for a  broad overview. The same SM parameters, which were so recently already assessed, were used for the new data analysis, too. The calculated cross sections have been obtained without as well as with the energy-dependent radius $r_D$ for the lower-energy surface imaginary potential of the well-deformed nuclei \cite{va14}. The additional data for $^{166}$Er show an even better agreement with the results corresponding to the energy-dependent $r_D$ quantity. A good agreement of the measured \cite{ggk15} and the final values of calculated cross sections for $^{164}$Er has also been obtained except the two lowest-energy points (Fig.~\ref{Fig:Er16246Re}). However, the agreement would exist even for these points if their energy-error bars would be similar to the next point at 14.51 MeV \cite{ggk15}.

On the other hand, the main conclusion of Ref. \cite{ggk15}, concerning the confirmed necessity of an energy-dependent modification of the $\alpha$-particle potential at very low energies, strengthens the firstly assumed \cite{ma09} and validated \cite{ma10,va14,va15} energy dependence of the surface imaginary-potential depth below as well as above 0.9$B$.

\subsubsection{$^{187}$Re} 

The target nucleus $^{187}$Re \cite{ps14} is quite interesting due to its place at the side of the well-deformed nuclei region. So, it has been of real interest to check in this case the assumption of an increased and energy-dependent $r_D$ radius for the well-deformed nuclei. Moreover, the corresponding $(\alpha,n)$ reaction cross sections have also been obtained with the code TALYS which presently includes the global potential \cite{va14} declared as the default option for the $\alpha$-particle optical potential \cite{TALYS}. 
The calculated cross sections correspond to the constant as well as increased and energy-dependent $r_D$ radii (Fig.~\ref{Fig:Er16246Re}). It is amazing how the first-time measured cross sections of this reaction \cite{ps14} are just in between the calculated cross sections related to the well-deformed and spherical nuclei, respectively, but closer to the former excitation function. 

One may note that the presently calculated data for $^{164,166}$Er and $^{187}$Re have been obtained by using the same $\alpha$-particle global potential \cite{va14}. On the contrary, the energy-dependence steepness assumed for the Fermi-type volume imaginary potential had the different parameter values of 2.5 \cite{ggk14} and 4 MeV \cite{ps14}, respectively.

Because the analysis of Ref. \cite{ps14} involved also the first of the three global OMPs by Demetriou {\it et al.} \cite{pd02}, which was identified as the $\alpha$-particle OMP that reproduces simultaneously the experimental data for all Ni isotopes \cite{as15} while the more complete and physical version is the third potential of the same authors, the results corresponding to all three potentials \cite{pd02} are also shown in Fig.~\ref{Fig:Er16246Re}. The larger underestimation of the measured data \cite{ps14} by the first potential of  Demetriou {\it et al.} is somehow reduced but not fully removed by using the other two potentials. 

\section{CONCLUSIONS} \label{Conc}

Recent high-precision $(\alpha,x)$ reaction data for the target nuclei $^{58,60-62,64}$Ni \cite{as15}, $^{107}$Ag \cite{cy15}, $^{106}$Cd \cite{ao15}, $^{112}$Sn \cite{ln15},  $^{164,166}$Er \cite{ggk15}, and $^{187}$Re \cite{ps14} have been analyzed in order to investigate the reliability of SM predictions on the basis of a previous $\alpha$-particle global potential \cite{va14}.
While the description of the $(\alpha,\gamma)$ as well as of the $(\alpha,n)$ and $(\alpha,p)$ reactions was found problematic within the above-mentioned original studies, a careful assessment of the related model parameters was firstly undertaken in the present work. This goal was achieved through various independent data analysis. Thus, the transmission coefficients of nucleons and $\gamma$ rays, given by the corresponding optical potentials and $\gamma$-ray strength functions, respectively, have particularly been fixed by independent analysis of $(p,n)$ reaction and radiative strength functions data. Then, they have also been checked by means of $(p,\gamma)$  reaction data. As a result, the accurate $(\alpha,x)$ reaction data became indeed uniquely sensitive to the $\alpha$-particle optical potential. 

There are several critical features of the statistical model parameters which led to particular conclusions of the present work. 
Thus, it has been found that the $\gamma$-ray energies larger than $\sim$3 MeV are more important for the calculation of the $(p,\gamma)$ reactions cross sections, while higher energies but yet below the nucleon binding energies play a more significant role for the $(\alpha,\gamma)$ reactions. Therefore, it results that an eventual low-energy RSF enhancement \cite{as14} would not affect essentially the calculated $(\alpha,\gamma)$ reaction cross sections. 

On the other hand, comparison with the results which have been obtained by using alternate parameters, particularly for $\alpha$-particle optical potential and RSFs, has shown that compensation effects of less suitable options of various SM parameters can not be avoided unless a consistent parameter set is used. 
Thus it has also been possible to point out that the real cause of the $(\alpha,p)$ reaction data underestimation below and around  the $(\alpha,n)$ reaction threshold, for the proton-rich nuclei $^{106}$Cd and $^{112}$Sn, has been not the $\alpha$-particle potential but a deficient RSF involved within the SM analysis. 
Actually, the largest importance of the RSF models for the suitable evaluation of the nucleon-capture cross sections has already been pointed out within, e.g., the detailed studies of Beard {\it et al.} \cite{mb12}. They have shown that the knowledge of RSFs is in that case even more crucial than information on nucleon OMPs.

The suitable description of all these recent reaction data has been proved possible with no empirical rescaling factors of the $\gamma$ and/or nucleon widths. It should be considered at once with the earlier data for nuclei in the whole range  45$\leq$$A$$\leq$209 which were involved within the assessment of the present optical potential \cite{va14,ma09,ma10}. Moreover, the agreement of the new measured and calculated $(\alpha,x)$ reaction cross sections proves the correctness of both the $\alpha$-particle global optical potential \cite{va14} itself and its set-up below $B$. Indeed, at these energies it was fully based on various SM calculations \cite{va14,ma09,ma10} which were carried out only by using consistent sets of SM parameters. 
Furthermore, its validation for the $\alpha$-particle emission in proton-induced reactions on Zn isotopes  \cite{va15} should be checked over the same $A$ range as for the incident channel in order to prove fully a global character.

\section*{Acknowledgments}

This work was partly supported by Fusion for Energy (F4E-GRT-168-02), and by Autoritatea Nationala pentru Cercetare Stiintifica (PN-42160102).

\end{document}